\documentclass[conference]{IEEEtran}
\IEEEoverridecommandlockouts
\usepackage{cite}
\usepackage{amsmath,amssymb,amsfonts}
\usepackage{algorithmic}
\usepackage{graphicx}
\usepackage{textcomp}
\usepackage{xcolor}

\usepackage{url}
\usepackage{color}
\usepackage{bm}
\usepackage{amsfonts}
\usepackage{amsmath}
\usepackage{booktabs}
\usepackage{threeparttable}
\usepackage{verbatim}

\def\BibTeX{{\rm B\kern-.05em{\sc i\kern-.025em b}\kern-.08em
    T\kern-.1667em\lower.7ex\hbox{E}\kern-.125emX}}
\begin{document}

\title{LiSenNet: Lightweight Sub-band and Dual-Path Modeling for Real-Time Speech Enhancement

\thanks{This work is supported by the National Natural Science Foundation of China (62101523), the joint AI laboratory of CMB-USTC (FTIT2022058), Hefei Municipal Natural Science Foundation (2022012) and USTC Research Funds of the Double First-Class Initiative (YD2100002008). (Correspondence: jzhang6@ustc.edu.cn)}
}

\author{\IEEEauthorblockN{Haoyin Yan\IEEEauthorrefmark{1}, Jie Zhang\IEEEauthorrefmark{1}, Cunhang Fan\IEEEauthorrefmark{2}, Yeping Zhou\IEEEauthorrefmark{3}, Peiqi Liu\IEEEauthorrefmark{3}}
\IEEEauthorblockA{\IEEEauthorrefmark{1}NERC-SLIP, University of Science and Technology of China (USTC), Hefei, China}
\IEEEauthorblockA{\IEEEauthorrefmark{2}Anhui Province Key Laboratory of Multimodal Cognitive Computation, Anhui University, Hefei, China}
\IEEEauthorblockA{\IEEEauthorrefmark{3}China Merchants Bank (CMB), Shenzhen, China}
}

\maketitle

\begin{abstract}
Speech enhancement (SE) aims to extract the clean waveform from noise-contaminated measurements to improve the speech quality and intelligibility. Although learning-based methods can perform much better than traditional counterparts, the large computational complexity and model size heavily limit the deployment on latency-sensitive and low-resource edge devices. In this work, we propose a lightweight SE network (LiSenNet) for real-time applications. We design sub-band downsampling and upsampling blocks and a dual-path recurrent module to capture band-aware features and time-frequency patterns, respectively. A noise detector is developed to detect noisy regions in order to perform SE adaptively and save computational costs. Compared to recent higher-resource-dependent baseline models, the proposed LiSenNet can achieve a competitive performance with only 37k parameters (half of the state-of-the-art model) and 56M multiply-accumulate (MAC) operations per second.
\end{abstract}

\begin{IEEEkeywords}
Speech enhancement, lightweight network, low complexity, real-time applications.
\end{IEEEkeywords}

\section{Introduction}
\label{sec:intro}
Speech enhancement (SE) aims to improve the instrumental quality and intelligibility of speech signals that are contaminated by ambient noises. Nowadays, speech-based applications are prevalent as a front-end module in systems, e.g., human-machine interaction, telephony, video-conferencing~\cite{chen15o_interspeech,7807946,8683385}. 
Traditional signal processing techniques are based on well-established theories and reasonable assumptions but often suffer from non-stationary acoustic scenes~\cite{zhang2023sdw}.
With the development of deep learning, deep neural network (DNN) based models can recover the clean speech in an end-to-end manner, even showing a superiority in the non-stationary case.

Formulated as a supervised learning problem, many DNN-based SE methods with impressive performance have been proposed recently. For example, convolution recurrent network (CRN)~\cite{tan18_interspeech} utilizes convolution encoder-decoder and recurrent neural network (RNN) to capture local features and model temporal patterns. Deep complex convolution recurrent network (DCCRN)~\cite{hu20g_interspeech} improves the CRN by complex-valued operations.
FullSubNet~\cite{9414177} is a cascade of the full-band and sub-band models, allowing to capture the global context while retaining the ability to model signal stationarity.

In spite of the promising performance, the large number of parameters and heavy computational burden hinder the deployment of these models on low-resource devices, e.g., hearing aids, headphones, IoT devices. This promotes the necessity of reducing the network complexity while maintaining performance. RNNoise~\cite{8547084} adopts a lightweight network to predict ideal critical band gains, and the coarse resolution of the bands is addressed by a pitch filter. 
In~\cite{DANG202332}, a two-stage model is proposed to reduce the model size and computational complexity, which combines coarse-grained full-band mask estimation and low-frequency refinement. In~\cite{10446016}, a lightweight model called FSPEN employs sub-band and full-band encoders to extract speech features. The FSPEN also utilizes an inter-frame path extension method to improve the dual-path architecture~\cite{9054266}. These methods can reach a balance between performance and complexity to some extent, yet it is still meaningful to develop a more lightweight SE model with comparable performance, which must be more appropriate for the deployment on low-resource edge devices.

In this work, we therefore propose a \textbf{Li}ghtweight \textbf{S}peech \textbf{en}hancement \textbf{Net}work,  called \textbf{LiSenNet} for real-time speech applications. A sub-band downsampling-upsampling method is developed to capture band-aware features, and a dual-path recurrent module is designed to capture time and frequency dependencies efficiently.
A  post-processing step is then utilized to refine the spectral phase to improve the perceptual quality.
In order to further reduce the computational burden, we apply a noise detector as an optional front-end module to detect noisy regions. This is rather benefitial for the inference efficiency in the instantaneous noise case. Experimental results show that our model achieves a competitive performance with only 37k parameters and 56M MACs per second. The noise detector can significantly reduce the real time factor (RTF) in case of low noise proportions in speech.
The reproducible code and audio examples are available online\footnote{https://github.com/hyyan2k/LiSenNet}.

\section{METHODOLOGY}
\label{sec:metho}

\begin{figure*}
    \centering
    \includegraphics[width=1.0\textwidth]{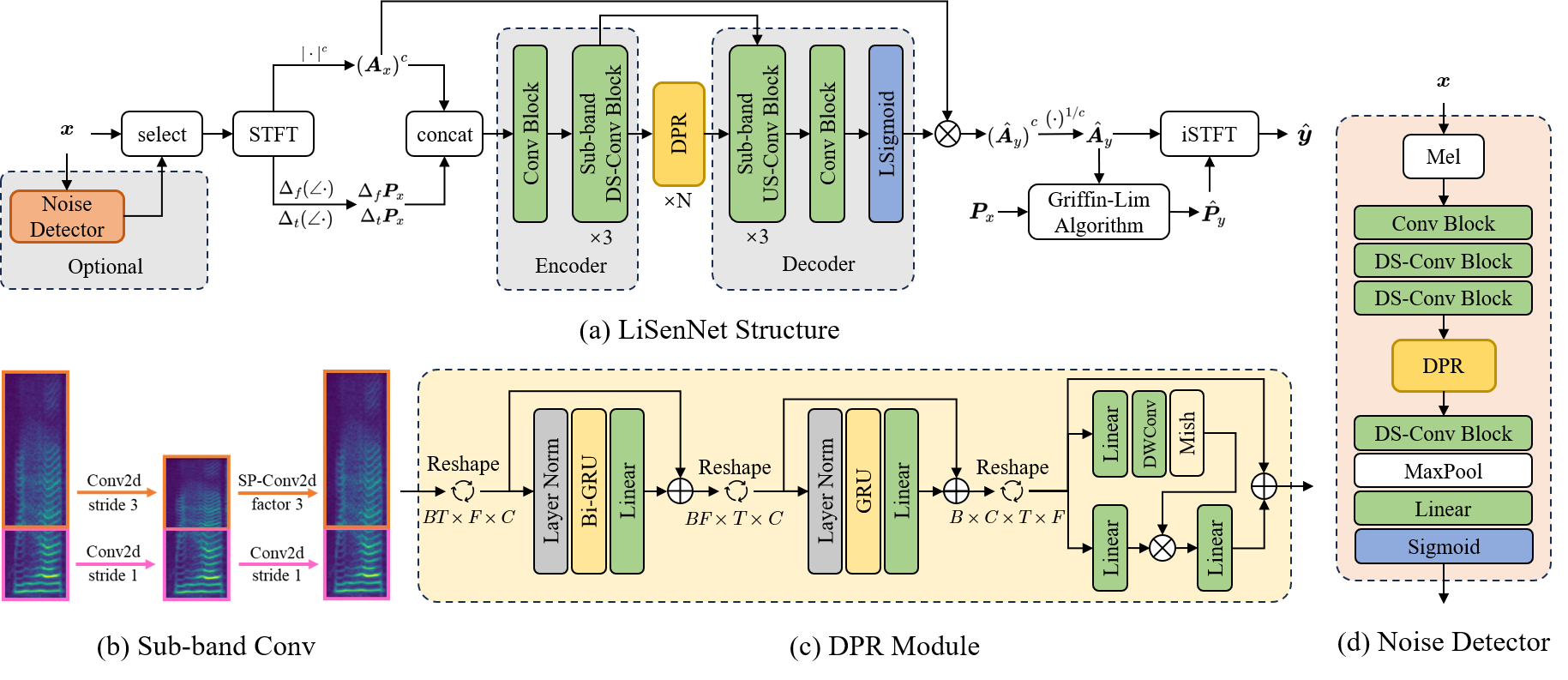}
    \caption{(a) The framework of our proposed LiSenNet, where $|\cdot|$ and $\angle\cdot$ denote the magnitude and phase extractors, $\Delta_f$ and $\Delta_t$ the differential operators along frequency and time axis, $(\cdot)^c$ and $(\cdot)^{1/c}$ the power compress and decompress operations at a ratio of $c$, respectively; (b) an example of sub-band DS-Conv and sub-band US-Conv; (c)  the proposed DPR module;  (d) the optional noise detector, where Mel spectrogram is extracted as the input feature.}
    \label{fig:architecture}
    \vspace{-1em}
\end{figure*}

An overview of the proposed LiSenNet is shown in Fig. \ref{fig:architecture}(a), which aims to recover the clean speech $\hat{\bm{y}} \in \mathbb{R}^L$ from the noisy input $\bm{x} \in \mathbb{R}^L$, where $L$ denotes the waveform length. Next, we will introduce the model in detail.

\subsection{Model Input}

First, we extract the complex spectrum $\bm{X} \in \mathbb{C}^{T \times F}$ of the noisy speech using the short-time Fourier transform (STFT) with $T$ and $F$ denoting the total time and frequency amounts, respectively. The power-compressed magnitude spectrum ${(\bm{A}_x)}^c \in \mathbb{R}^{T \times F}$ and phase spectrum $\bm{P}_x \in \mathbb{R}^{T \times F}$ can then be calculated, where $c$ is the compression ratio. Similarly to~\cite{9522648}, we set $c$ = 0.3 to equalize the importance of quieter sounds relative to loud ones. 

Since phase difference (PD) exhibits similar patterns as the magnitude spectrum~\cite{6954314}, we also adopt the PD as input to help estimate the clean magnitude spectrum.
The time-frequency (TF) domain PDs  are computed  as
\begin{align}
\Delta_f \bm{P}_x(t,f) =& \bm{P}_x(t,f) - \bm{P}_x(t,f-1), \\
\Delta_t \bm{P}_x(t,f) =& \bm{P}_x(t,f) - \bm{P}_x(t-1,f) - 2\pi f {Q \over M}, \label{Dt}
\end{align}
where $t$ and $f$ are the frame and frequency indices, $Q$ and $M$ the frame shift and frame length, respectively.
As the STFT is performed on overlapped frames, resulting in an offset in the phase, which would distort the structure of temporal PD, we thus subtract $2\pi f {Q \over M}$ in \eqref{Dt} to obtain the baseband PD~\cite{6891278}. 
The power-compressed magnitude spectrum and noisy PD are concatenated as the model input.

\subsection{Encoder and Decoder}
In Fig. \ref{fig:architecture}(a), both the encoder and decoder contain a series of convolution (Conv) blocks, where each block consists of a convolution layer, a layer normalization and a parametric rectified linear unit (PReLU) activation. To reduce the computational  complexity, we exploit downsampling convolution (DS-Conv) blocks to halve the frequency-axis size in the encoder and upsampling convolution (US-Conv) blocks to recover the frequency resolution in the decoder. Skip connection is conducive to integrating low-level and high-level features.

Due to the fact that in general low-frequency bands
play a more important role in human hearing than high-frequency bands~\cite{Hermansky1990PerceptualLP}, we propose the sub-band DS-Conv and sub-band US-Conv to maintain low-frequency resolution, e.g. see the illustrative example in Fig.~\ref{fig:architecture}(b). For low-frequency bands, Conv2d with a stride of 1 is utilized. For high-frequency bands, we apply Conv2d with a stride of 3 in the downsampling and sub-pixel convolution (SP-Conv2d)~\cite{7780576} with factor 3 in the upsampling. Therefore, the overall downsampling and upsampling factor of frequency bands becomes 2.

The learnable Sigmoid (LSigmoid)~\cite{fu21_interspeech} is adopted as the last layer of the decoder to predict the mask, given by
\begin{align}
{\rm LSigmoid}(x) = \beta /\left(1 + e^{-\alpha x}\right),
\end{align}
where $\beta$ is a constant scalar set to 2.0, and $\alpha \in \mathbb{R}^{F}$ is trainable. Considering distinct patterns in high and low frequency bands of speech signals, LSigmoid can adaptively learn the function shapes for different bands.

\subsection{Dual-Path Recurrent (DPR) Module}

Since the emergence of DPRNN~\cite{9054266}, dual-path dependent models have shown to be very promising for SE~\cite{10446016, le21b_interspeech, cao22_interspeech}. Specifically, time and frequency sequences are modeled sequentially to capture the time dependency in each band and the frequency dependency in each frame. In this work, we propose a DPR module as the backbone in Fig. \ref{fig:architecture}(c), where the feature map $\bm{M} \in \mathbb{R}^{B \times C \times T \times F}$ is reshaped to $BT \times F \times C$ for frequency modeling and then reshaped to $BF \times T \times C$ for temporal modeling with $B$ and $C$ being the batch size and hidden dimension, respectively.
To ensure the causality of our model, we apply bidirectional GRU (Bi-GRU) for frequency modeling and unidirectional GRU for temporal modeling. The RNN nodes need to be decreased to reduce the computational complexity for real-time applications, resulting in the decay of modeling capacity. To address this issue, we utilize the convolutional gated linear unit (ConvGLU) modified from \cite{shi2024transnext} as a channel mixer to fuse inter-channel information, which is composed of a linear layer, depthwise convolution \cite{8099678} (DWConv) and Mish \cite{misra2019mish} activation. 
The DWConv can aggregate the nearest information and the gate mechanism allows for a  fine-grained channel attention.

\subsection{Phase Refinement}
Instead of directly using the noisy phase $\bm{P}_x$ to synthesize the output speech~\cite{tan18_interspeech,pmlr-v97-fu19b}, we propose to utilize the Griffin-Lim Algorithm (GLA)~\cite{1172092, 6701851} as a post-processor to refine the phase spectrum. The GLA performs phase retrieval depending on the spectral consistency, which iteratively applies STFT and inverse STFT (iSTFT) to update the phase as
\begin{align}
\bm{P}^{(k)} = \angle {\rm STFT}\left({\rm iSTFT}\left(\bm{A}e^{i\bm{P}^{(k-1)}}\right)\right),
\end{align}
where $k$ denotes the iteration number and $\bm{A}$ denotes the given magnitude. The GLA has been considered for blind source separation, e.g., in~\cite{6287827}. However, errors might be introduced at each iteration due to the possible speech distortion and residual noise in the enhanced magnitude spectrum, resulting in a limited phase accuracy~\cite{10.1121/1.4986647}.
Since advanced SE models can generate a more accurate magnitude, we propose to reconsider GLA for phase refinement as
\begin{align}
\bm{\hat{P}}_y={\rm GLA}\left(\hat{\bm{A}}_y, \bm{P}_x\right),
\end{align}
where $\hat{\bm{A}}_y$ denotes the estimated magnitude spectrum.

\subsection{Noise Detector (ND)}

The efficiency of existing SE models might decrease in the case of instantaneous noises, where the noise duration can be much shorter than the speech length. Turning on the SE module continuously would cause a large amount of unnecessary resource consumption. Therefore, as depicted in Fig. \ref{fig:architecture}(d), we propose a noise detector as an optional front-end to detect frame-level noisy regions. We extract 64-dimensional Mel spectrograms to reduce the redundancy of the input features. The output binary values of the Sigmoid layer indicate whether each frame contains noise (0 for clean and 1 for noisy).
Only noisy frames need to be sent to the subsequent SE back-end, such that the overall computational burden can be further saved, particularly in short noise cases.

\subsection{Loss Function}

In this work, we consider multiple loss functions to train the proposed LiSenNet. Similarly to \cite{cao22_interspeech}, we apply the mean squared error (MSE) on the power-compressed magnitude spectrum and complex spectrum as
\begin{align}
    \mathcal{L}_{\rm mag}&=\mathbb{E}\left[{|| {(\bm{A}_y)}^c - {(\hat{\bm{A}}_y)}^c||^2}\right], \\
    \mathcal{L}_{\rm comp}&=\mathbb{E}\left[{|| {\bm{Y}}^c - {\hat{\bm{Y}}}^c||^2}\right],
\end{align}
where ${\bm{Y}}^c$ = ${(\bm{A}_y)}^c e^{i\bm{P}_y}$ and ${\hat{\bm{Y}}}^c$ = ${(\hat{\bm{A}}_y)}^c e^{i\hat{\bm{P}}_y}$ denote the power-compressed complex spectrum of the clean and estimated speech, respectively. 

Following \cite{pmlr-v97-fu19b}, we use a discriminator $D$ to further improve the perceptual speech quality, which is trained to predict the perceptual evaluation of speech quality (PESQ)~\cite{941023}, which is scaled to (0, 1).
The discriminator encourages the model to generate the magnitude with a larger PESQ score, and the PESQ loss is given by
\begin{align}
    \mathcal{L}_{\rm pesq} = \mathbb{E}\left[|| D({(\bm{A}_y)}^c, {(\hat{\bm{A}}_y)}^c) -1 ||^2\right].
\end{align}

If the noise detector is included, we choose the binary cross entropy (BCE) loss $\mathcal{L}_{\rm bce}$ as the training criterion. Therefore, the overall loss function for model training is given by
\begin{align}
    \mathcal{L} = \lambda_1 \mathcal{L}_{\rm mag} + \lambda_2 \mathcal{L}_{\rm comp} + \lambda_3 \mathcal{L}_{\rm pesq} + \mathcal{L}_{\rm bce},
\end{align}
where $\lambda_1,\lambda_2,\lambda_3$ are the empirical weights, which are respectively set to 0.9, 0.1 and 0.05 in this work.

\section{EXPERIMENTS}
\label{sec:exp}

\subsection{Experimental Setup}

We use two datasets to evaluate our proposed method. The first one is the widely-used SE benchmark VoiceBank+DEMAND~\cite{valentinibotinhao16_ssw}, which provides clean-noisy audio pairs. There are 11,572 utterances from 28 speakers for training and 872 utterances from 2 speakers for testing. 
Clean signals are mixed with noises at signal-to-noise ratios (SNRs) of \{0, 5, 10, 15\} dB in the training set and \{2.5, 7.5, 12.5, 17.5\} dB in the test set.
All clips are resampled from 48 kHz to 16 kHz in experiments.

In addition, we create the WSJ0+ESC50 dataset for the evaluation of the noise detector. The clean speech utterances are selected from the Wall Street Journal (WSJ0) dataset~\cite{wsj0}, and the noise sources are from ESC50~\cite{piczak2015dataset}, which includes 2000 environmental audio recordings. Each 4-second clip contains 0.5 to 2 seconds of noise, and the SNR is sampled uniformly between -10 and 5 dB.

We perform STFT  using a Hanning window with a length of 512 (32ms) and a hop size of 256 (16ms). The numbers of output channels for Conv blocks in the encoder, decoder, and noise detector are \{4, 8, 12, 16\}, \{12, 8, 4, 1\}, and \{4, 8, 16, 32\}, respectively. The number of hidden states is 24 for unidirectional GRU and 12 for Bi-GRU. The DPR module is repeated 2 times (i.e., $N$ = 2 in Fig.~\ref{fig:architecture}(a)) in the network and the iterations for GLA are set to 2. Our model is trained using the AdamW \cite{loshchilov2019decoupled} optimizer with 100 epochs. The $L_2$ norm for gradient clipping is set to 5.0. The learning rate starts from 5e-4 and reduces at a decay factor of 0.98 in each epoch.

\subsection{Experimental Results}

\begin{table}[]
    \centering
    \caption{Performance comparision on VoiceBank+DEMAND dataset.}
    \resizebox{\columnwidth}{!}{
        \begin{threeparttable}
        \begin{tabular}{lcccc}
            \toprule
             Method & Para. & MACs & PESQ & STOI \\
             \midrule
             Noisy & - & - & 1.97 & 0.921 \\
             RNNoise~\cite{8547084} & 0.06M & \textbf{0.04G} & 2.33 & 0.922 \\
             {CRN\tnote{*} }~\cite{tan18_interspeech} & 17.58M & 2.56G & 2.54 & 0.936 \\
             {DCCRN\tnote{*} }~\cite{hu20g_interspeech} & 3.67M & 14.38G & 2.78 & 0.940 \\
             {FullSubNet\tnote{*} }~\cite{9414177} & 5.64M & 30.87G & 2.89 & 0.940 \\
             {Fast FullSubNet\tnote{*} }~\cite{hao2023fastfullsubnet} & 6.84M & 4.14G & 2.79 & 0.935 \\
             CCFNet+ (Lite)~\cite{DANG202332} & 160k & 390M & 2.94 & - \\
             FSPEN~\cite{10446016} & 79k & 89M & 2.97 & \textbf{0.942} \\
             \midrule
             LiSenNet & \textbf{37k} & 56M & \textbf{3.07} & 0.939 \\
             \bottomrule
        \end{tabular}
    
        \begin{tablenotes}
        \footnotesize
        \item[*] We retrained these models to obtain results.
        \end{tablenotes}
        
        \end{threeparttable}
    }
    \vspace{-0.5em}
    \label{tab:comparison1}
\end{table}

\begin{table}[]
    \centering
    \caption{Performance comparision on WSJ0+ESC50 dataset.}
    \resizebox{\columnwidth}{!}{
        \begin{tabular}{lcccc}
            \toprule
             Method & Para. & MACs & RTF & PESQ  \\
             \midrule
             Noisy & - & - & - & 1.86 \\
             CRN~\cite{tan18_interspeech} & 17.58M & 2.56G & 0.042 & 2.80 \\
             DCCRN~\cite{hu20g_interspeech} & 3.67M & 14.38G & 0.184 & 3.05 \\
             FullSubNet~\cite{9414177} & 5.64M & 30.87G & 0.386 & 2.99 \\
             Fast FullSubNet~\cite{hao2023fastfullsubnet} & 6.84M & 4.14G & 0.100 & 2.75 \\
             \midrule
             LiSenNet & \textbf{37k} & 56M & 0.028 & \textbf{3.08} \\
             LiSenNet+ND & 55k & \textbf{15M$\sim$70M} & \textbf{0.020} & 3.00 \\
    
             \bottomrule
        \end{tabular}
    }
    \vspace{-0.5em}
    \label{tab:comparison2}
\end{table}

First, we show the performance of the proposed LiSenNet and other state-of-the-art (SOTA) methods in terms of the parameter amount (Para.), MACs, PESQ, and STOI~\cite{5713237} on the VoiceBank+DEMAND dataset in Table \ref{tab:comparison1}. It is clear that LiSenNet achieves competitive performance with a tiny model size and low computational burden.
Compared to RNNoise which has a comparable resource consumption, our model shows an obvious superiority in the SE performance.
In comparison with FSPEN, our LiSenNet improves the PESQ score by 0.1 with only 47\% parameters and 63\% MACs.

\begin{figure}
    \centering
    \includegraphics[width=0.92\columnwidth]{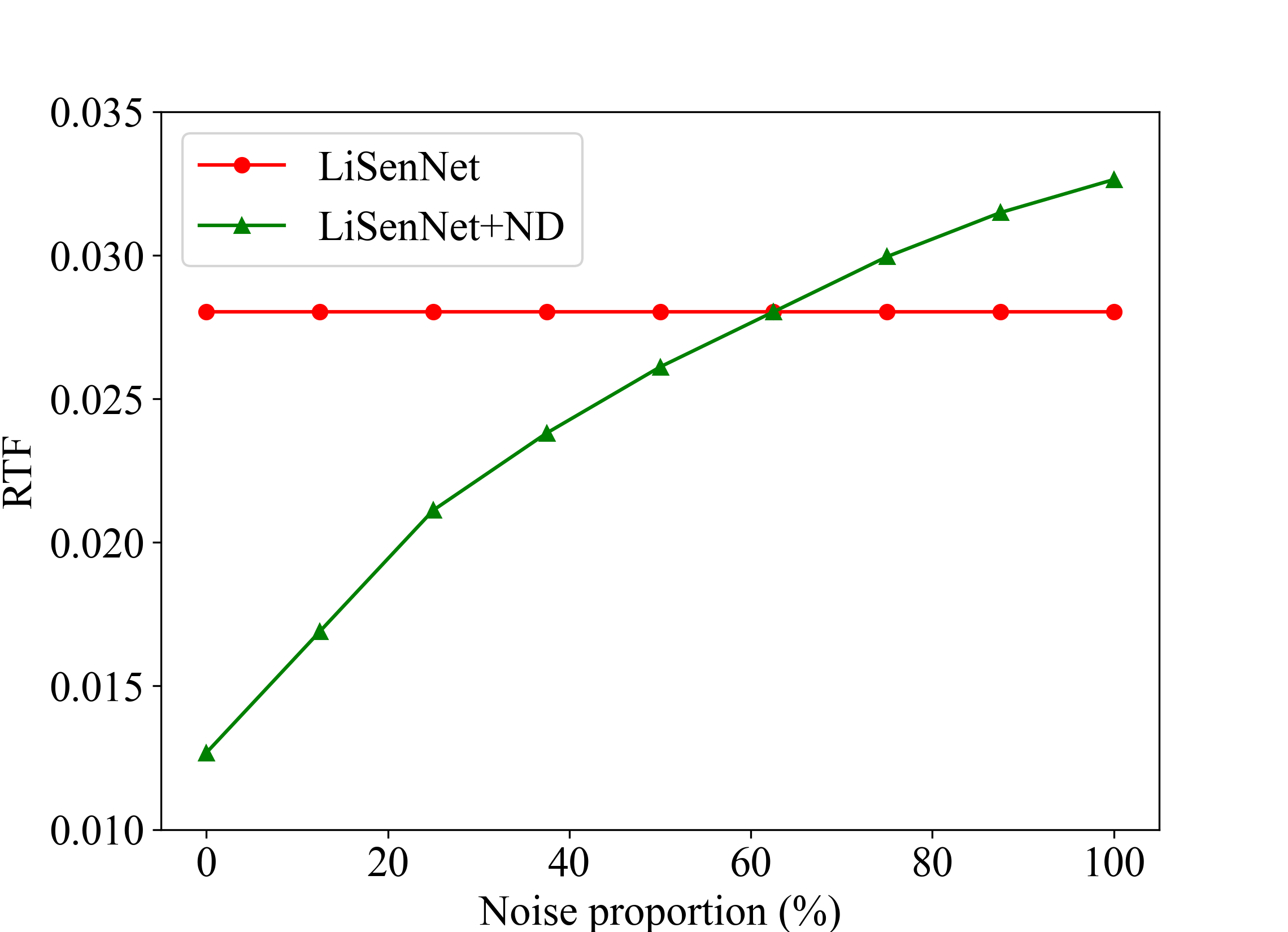}
    \caption{The RTF under different noise proportion conditions.}
    \vspace{-0.5em}
    \label{fig:rtf}
\end{figure}

Results on the WSJ0+ESC50 dataset are presented in Table~\ref{tab:comparison2}. The RTF is measured with Intel(R) Xeon(R) E5-2620 v4 CPU. LiSenNet obtains the best performance with a much lower complexity in comparison with SOTA methods. At the expense of a slight performance decay, which is caused by the imprecision of noise detection and the information loss of clean regions, the inclusion of noise detector (ND) can further reduce the computational complexity. But this (of LiSenNet+ND) depends on the noise proportion, ranging from 15M to 70M in MACs, which however is still much lower than other methods. In the extreme case (no noise in speech), the proposed method only requires 15M MACs/second. 
Fig.~\ref{fig:rtf} shows the RTF curve in terms of the noise proportion, demonstrating the effectiveness of the noise detector as a front-end, especially when noise proportion in speech is low.

Finally, we carry out an ablation study to analyze the impact of different components of the proposed LiSenNet in Table \ref{tab:ablation}.
The ablation of input features and loss function validates that PD helps the magnitude estimation and $\mathcal{L}_{\rm pesq}$ is beneficial for the perceptual quality.
Replacing the sub-band Conv with a normal convolution layer (w/o sub-band Conv), removing the ConvGLU module (w/o ConvGLU), or freezing the learnable parameters in LSigmoid (w/o LSigmoid) will lead to a decrease in performance, verifying the necessity of these modules.
The number of the DPR module can make a trade-off between complexity and performance, and $N$ = 2 seems the best choice.
Furthermore, the GLA is effective for refining the phase spectrum, but 2 iterations ($k$ = 2) are enough for the convergence of performance.

\begin{table}[]
    \centering
    \caption{Ablation study on VoiceBank+DEMAND dataset.}
    \resizebox{\columnwidth}{!}{
        \begin{tabular}{lcccc}
            \toprule
             Method & Para. & MACs & PESQ & STOI \\
             \midrule
             LiSenNet & 37k & 56M & \textbf{3.07} & 0.939 \\
             \midrule
             w/o PD & 37k & 56M & 2.99 & 0.936 \\
             w/o $\mathcal{L}_{\rm pesq}$ & 37k & 56M & 2.95 & 0.937 \\
             \midrule
             w/o sub-band Conv & 30k & 54M & 3.01 & 0.938 \\
             w/o ConvGLU & 31k & 48M & 2.98 & 0.937 \\
             w/o LSigmoid & 37k & 56M & 3.05 & 0.938 \\
             \midrule
             DPR ($N$ = 0) & \textbf{15k} & \textbf{22M} & 2.68 & 0.913 \\
             DPR ($N$ = 1) & 26k & 39M & 2.97 & 0.935 \\
             DPR ($N$ = 2) & 37k & 56M & 3.07 & 0.939 \\
             DPR ($N$ = 3) & 48k & 72M & 3.06 & \textbf{0.940} \\
             \midrule
             GLA (Noisy Phase) & 37k & 56M & 3.02 & 0.939 \\
             GLA ($k$ = 1)  & 37k & 56M & 3.06 & 0.939 \\
             GLA ($k$ = 2) & 37k & 56M & 3.07 & 0.939 \\
             GLA ($k$ = 3) & 37k & 56M & 3.07 & 0.939 \\
             \bottomrule
        \end{tabular}
    }
    \vspace{-0.5em}
    \label{tab:ablation}
\end{table}

\section{CONCLUSION}
\label{sec:conc}

In this work, we proposed a lightweight network called LiSenNet for efficient real-time SE. The sub-band 
DS-Conv and US-Conv were proposed to maintain the resolution of low-frequency bands, and the dual-path recurrent module was used to model the intra-frame, inter-frame and inter-channel patterns. The denoised magnitude spectrum was estimated using the noisy magnitude spectrum and phase difference, and the GLA was adopted to refine the phase spectrum. The noise detector can be flexibly inserted  to further reduce the computational burden depending on the noise proportion. The proposed LiSenNet showed a superiority in both the SE performance and complexity, which is thus more appropriate for the deployment onto low-resource edge devices.

\vfill\pagebreak

\bibliographystyle{ieeetr}
\bibliography{refs}

\end{document}